\newcommand*{\ATLASLATEXPATH}{atlaslatex/latex/}
\documentclass[UKenglish,texlive=2011,cernpreprint,txfonts,PAPER]{\ATLASLATEXPATH atlasdoc}
\usepackage[biblatex=true,backend=biber,full,subfigure]{\ATLASLATEXPATH atlaspackage}
\usepackage{\ATLASLATEXPATH atlasbiblatex}

\usepackage{\ATLASLATEXPATH atlascontribute}
\usepackage{authblk}

\usepackage{\ATLASLATEXPATH atlasphysics}

\graphicspath{{atlaslatex/logos/}{figures/}{./}}

\usepackage{atlas-document-defs}

\AtlasTitle{Measurement of the charge asymmetry in highly boosted top-quark pair production in $\sqrt{s} =$ 8~$\mathrm{\ Te\kern -0.1em V}${} $pp$ collision data collected by the ATLAS experiment}

\author{The ATLAS Collaboration}

\AtlasRefCode{TOPQ-2015-10}

\PreprintIdNumber{CERN-PH-EP-2015-303}

\AtlasJournalRef{\PLB 756 (2016) 52-71}
\AtlasDOI{10.1016/j.physletb.2016.02.055}

\AtlasAbstract{%
 In the $pp \rightarrow t\bar{t}$ process the angular distributions of 
top and anti-top quarks are expected to present a subtle difference, which 
could be enhanced by processes not included in the Standard Model. This Letter presents a measurement  
of the charge asymmetry in events where the top-quark pair is produced with 
a large invariant mass. The analysis is performed on 20.3~\mbox{fb$^{-1}$}{} of $pp$ collision data 
at $\sqrt{s} =$ 8~$\mathrm{\ Te\kern -0.1em V}${} collected by the ATLAS 
experiment at the LHC, using reconstruction techniques specifically designed 
for the decay topology of highly boosted top quarks. The charge asymmetry 
in a fiducial region with large invariant mass of the top-quark pair 
($m_{\ttbar} > $ 0.75~$\mathrm{\ Te\kern -0.1em V}$) and an absolute rapidity difference of the top and
anti-top quark candidates within $-$2 $ < |y_t| - |y_{\bar{t}}| <$ 2 is 
measured to be 4.2 $\pm$ 3.2\%, 
in agreement with the Standard Model prediction at next-to-leading order. A differential measurement in three $t\bar{t}${} mass bins is also presented.
}

\hypersetup{pdftitle={ATLAS draft},pdfauthor={The ATLAS Collaboration}}

\begin{document}

\maketitle

\section{Introduction}
\label{sec:intro}

The charge asymmetry~\cite{Kuhn:1998kw,Kuhn:1998jr} in top-quark pair production at hadron colliders constitutes
one of the more interesting developments in the last decade of top-quark physics. 
In the Standard Model (SM), a forward--backward asymmetry ($A_{\mathrm{FB}}$), of 
order $\alpha_s$, is expected at a proton--antiproton ($p\bar{p}$) collider such as the Tevatron, with a much enhanced asymmetry 
in certain kinematical regions. Early measurements~\cite{Abazov:2007ab,Aaltonen:2008hc} found a larger $A_{\mathrm{FB}}$ than predicted by the SM. Later determinations confirmed this deviation 
and measurements in intervals of the invariant mass, $m_{\ttbar}$, of the system formed by the top-quark 
pair~\cite{Abazov:2015fna,Abazov:2014cca,Aaltonen:2012it,Abazov:2011rq,Aaltonen:2011kc} found a stronger dependence on $m_{\ttbar}$ than anticipated. 
Recent calculations of electroweak effects~\cite{Hollik:2011ps} and the full next-to-next-to-leading-order (NNLO) 
corrections~\cite{Czakon:2014xsa} to the asymmetry have brought the difference between the observed asymmetry at the Tevatron and the SM prediction down to the 1.5 
$\sigma$ level and reduced the tension with the differential measurements in $m_{\ttbar}$~\cite{Aguilar-Saavedra:2014kpa,Kuhn:2011ri}. 

At the Large Hadron Collider (LHC), the forward--backward asymmetry is not present due to the symmetric initial state, but a related charge asymmetry, $\AC{}$, is expected in the 
distribution of the difference of absolute rapidities of the top and anti-top quarks,

\begin{equation}
\label{eq:ac}
\AC{}= \frac{N(\dy{} >0) - N(\dy{} <0)}
       {N(\dy{} >0) + N(\dy{} <0)},
\end{equation}

where $\dy{} = |y_{t}|-|y_{\bar{t}}|$ and $y$ denotes the rapidity of the top and anti-top quarks.
\footnote{ATLAS uses a right-handed coordinate system with its origin at the nominal interaction 
point (IP) in the centre of the detector and the $z$-axis coinciding with the axis of the beam pipe. 
The $x$-axis points from the IP towards the centre of the LHC ring, and the $y$-axis points upward. 
Polar coordinates ($r$, $\phi$) are used in the transverse plane, $\phi$ being the azimuthal angle 
around the $z$-axis. The rapidity $y$ is given as $y = - \frac{1}{2} \ln[(E + p_z)/(E - p_z)]$, 
while the pseudorapidity is defined in terms of the polar angle $\theta$ as $\eta = - \ln[{\tan{(\theta/2)}}]$. 
The distance in ($\eta$,$\phi$) coordinates, $\Delta R = \sqrt{(\Delta \phi)^2 + (\Delta \eta)^2}$, is used 
to define cone sizes and the distance between reconstructed objects. Transverse momentum and energy are 
defined as $p_{\mathrm{T}} = p \sin \theta$ and $E_{\mathrm{T}} = E \sin \theta$, respectively.} For quark--antiquark ($q\bar{q}$) initial states, 
the difference in the average momentum carried by valence and sea quarks leads to a positive asymmetry. 
These quark-initiated processes are strongly diluted by the charge-symmetric gluon-initiated processes, 
yielding a SM expectation for the charge asymmetry of less than 1\%. Many beyond-the-Standard-Model (BSM) 
scenarios predict an alteration to this asymmetry. Previous measurements at 7~\tev~\cite{Chatrchyan:2014yta,Aad:2013cea,Chatrchyan:2012cxa,ATLAS:2012an} 
and 8~\tev~\cite{Khachatryan:2015oga,Khachatryan:2015mna,ATLAS-TOPQ-2014-16} by ATLAS and CMS are consistent with the SM prediction. 

With a centre-of-mass energy of 8~\tev{} and a top-quark pair sample
of millions of events, the LHC experiments can access the 
charge asymmetry in a kinematic regime not probed by previous experiments. 
The development of new techniques involving Lorentz-boosted objects and jet substructure~\cite{Adams:2015hiv,Altheimer:2013yza,Altheimer:2012mn,Abdesselam:2010pt} 
and their use in the analysis of LHC data~\cite{Aad:2013gja,ATLAS:2012am} have enabled an efficient 
selection of highly boosted objects and an accurate reconstruction 
of their momentum. 

This Letter presents a measurement of the rapidity-dependent charge
asymmetry in top-quark pair production that is based on techniques specifically 
designed to deal with the collimated decay topology of boosted top quarks. Specifically, it is based on the 
techniques described in Refs.~\cite{Aad:2015fna,Aad:2013nca,Aad:2012dpa,ATL-PHYS-PUB-2010-008}. 
The analysis focuses on the lepton+jets 
($\ell$+jets) final state, where the hadronic top-quark decay is 
reconstructed as a single large-radius (large-$R$) jet and tagged as such 
using jet substructure variables. 
The leptonic top-quark decay is reconstructed from a single small-radius 
(small-$R$) jet, a single 
charged lepton (muon or electron), and missing transverse momentum, corresponding to the neutrino from the $W$ boson decay. The event
selection and reconstruction follow the prescriptions of Ref.~\cite{Aad:2015fna},
where a detailed description and discussion of their performance can be found.

Compared to previous analyses~\cite{Khachatryan:2015oga,ATLAS-TOPQ-2014-16} 
based on the classical, {\em resolved} top-quark selection criteria and 
reconstruction schemes, 
this approach offers a more precise reconstruction of the \ttbar{} 
invariant mass and top-quark direction for highly boosted top quarks. It is  
therefore possible to perform accurate measurements of the charge asymmetry
in events with a \ttbar{} invariant mass in the \tev{} range. This kinematic 
regime has a higher sensitivity for the SM asymmetry due to a higher 
fraction of quark-initiated processes, as well as for BSM models that 
introduce massive new states. 

This Letter is structured as follows. 
The data sample analysed is presented in Section~\ref{sec:data}, along with a description
of the Monte Carlo (MC) simulation samples in Section~\ref{sec:mc}. 
A brief overview of the reconstructed object 
definitions and of the event selection and reconstruction is given in 
Sections~\ref{sec:obj} and~\ref{sec:evtselreco}.
The observed yields and several kinematic distributions are compared to the SM expectations
in Section~\ref{sec:yield}. The unfolding technique used to correct the 
reconstructed \dy{} spectrum to the parton level is discussed 
in Section~\ref{sec:unfolding}. The estimates of the systematic uncertainties 
that affect the measurement are described and estimated in Section~\ref{sec:syst}. The results 
are presented in Section~\ref{sec:results}, and their impact on several
BSM theories is discussed in Section~\ref{sec:bsm}.
Finally, the conclusions are presented in Section~\ref{sec:conclusions}.

\section{Data sample}
\label{sec:data}

The data for this analysis were collected by the ATLAS~\cite{Aad:2008zzm}
experiment in the 8~\tev{} proton--proton ($pp$) collisions at the CERN LHC in 2012. 
Collision events are selected 
using isolated or non-isolated single-lepton triggers, where the isolated triggers have a threshold of 24~\gev{} on the transverse
momentum (\pt) of muons or on the transverse energy
of electrons. The non-isolated triggers have higher thresholds: 
60~\gev{} for electrons and 36~\gev{} for muons. The contribution from events with leptons passing only the non-isolated triggers but having $\pt$ below these higher thresholds is negligible. The collected data set is limited to periods with stable beam conditions when all sub-systems were operational. The sample corresponds to an 
integrated luminosity of 20.3 $\pm$ 0.6~\ifb{}.

\section{Monte Carlo simulation}
\label{sec:mc}
Samples of MC simulated events are used to characterise the detector
response and efficiency to reconstruct $\ttbar$ events, estimate systematic 
uncertainties, and predict the background contributions from various 
physics processes. The response of the detector and trigger is simulated~\cite{ATLASSim} 
using a detailed model implemented in GEANT4~\cite{Agostinelli:2002hh}.
Simulated events are reconstructed with the same software
as the data. Additional $pp$ interactions, simultaneously present in 
the detector (pile-up), are generated using {\normalfont \scshape Pythia} 
8.1~\cite{Sjostrand:2007gs}
with the MSTW2008 leading order PDF set~\cite{mstw} and the AUET2 set of tune
parameters (tune). The
pile-up events are reweighted to the number of interactions per bunch 
crossing in data (on average 21 in 2012). 
For some samples used to evaluate systematic uncertainties,
the detailed description of the calorimeter response is parameterised using the
ATLFAST-II simulation~\cite{ATLASSim}. For all samples the top-quark mass is set to $m_{\mathrm{top}} =$ 172.5~\GeV.

The nominal signal $\ttbar$ sample is produced 
using the {\normalfont \scshape Powheg-Box} (version 1, r2330) 
generator~\cite{Frixione:2007vw}, which is based on next-to-leading-order 
(NLO) QCD matrix elements.
The CT10 \cite{CT10PDF} set of parton distribution
functions (PDF) is used. The $h_{\mathrm{damp}}$ parameter, which controls the matrix element
(ME) to parton shower (PS) matching in {\normalfont \scshape Powheg-Box} 
and effectively regulates the high-\pt{} radiation, 
is set to the top-quark mass.
The parton shower, hadronisation, and the underlying event are simulated 
with {\normalfont \scshape Pythia} 6.427~\cite{SAMPLES-PYTHIA} using the CTEQ6L1 PDF set and the Perugia 2011~\cite{Skands} tune. 
Electroweak corrections are applied to this sample through a reweighting scheme; they are calculated with
{\normalfont \scshape Hathor 2.1-alpha}~\cite{Aliev:2010zk} implementing the theoretical calculations of Refs.~\cite{Kuhn:2005it,Kuhn:2006vh,Kuhn:2013zoa}. 
Alternative samples are used to evaluate uncertainties in modelling
the \ttbar{} signal. These include samples produced with
{\normalfont \scshape MC@NLO} 4.01\cite{SAMPLES-MCNLO} 
interfaced with {\normalfont \scshape Herwig} 6.520\cite{SAMPLES-HERWIG} 
and {\normalfont \scshape Jimmy} 4.31 \cite{SAMPLES-JIMMY}, as well as samples
generated with {\normalfont \scshape Powheg-Box} + {\normalfont \scshape Herwig}/{\normalfont \scshape Jimmy}
and {\normalfont \scshape Powheg-Box} + {\normalfont \scshape Pythia}, both
with $h_{\mathrm{damp}} = $ infinity. Samples are also produced with differing initial- and final-state 
radiation (ISR/FSR), using the {\normalfont \scshape AcerMC} generator~\cite{AcerMC35} interfaced with {\normalfont \scshape Pythia}. 
All $\ttbar$ samples are normalised to cross-section at NNLO + next-to-next-to-leading logarithmic (NNLL) accuracy\footnote{
  The top++2.0 \cite{Czakon:2011xx} 
  calculation includes the NNLO QCD
  corrections and resums NNLL soft gluon
  terms. The quoted cross-section corresponds to a top-quark mass of
  172.5~\GeV.} 
\cite{Cacciari:2011hy,Beneke:2011mq,Baernreuther:2012ws,Czakon:2012zr,Czakon:2012pz,Czakon:2013goa}: $\sigma_{t\bar{t}}=253^{+13}_{-15}$~pb.

Leptonic decays of vector bosons produced in association with several 
high-$\pt$ jets, referred to as $W$+jets and $Z$+jets events, 
with up to five additional final-state partons in the leading-order (LO)
matrix-elements, are produced with the {\normalfont \scshape Alpgen} generator~\cite{SAMPLES-ALPGEN} 
interfaced with {\normalfont \scshape Pythia} 6.426 for parton fragmentation using the MLM matching
scheme~\cite{Mangano:2001xp}. Heavy-flavour quarks
are included in the ME calculations to model the $Wb\bar{b}$,
$Wc\bar{c}$, $Wc$, $Zb\bar{b}$ and $Zc\bar{c}$ processes.
The $W$+jets samples are normalised 
to the inclusive $W$ boson NNLO cross-section \cite{Hamberg:1990np,Gavin:2012sy}. 
 
Single top-quark production is simulated using {\normalfont \scshape Powheg-Box} interfaced with 
{\normalfont \scshape Pythia} 6.425 using the CTEQ6L1 PDF set and the Perugia 2011 tune. 
The cross-sections multiplied by the sum of the branching ratios for the leptonic $W$ decay employed for these processes are 28 pb ($t$-channel) \cite{Kidonakis:2011wy},
22 pb ($Wt$ production) \cite{Kidonakis:2010ux}, and 1.8 pb ($s$-channel) \cite{Kidonakis:2010tc}, obtained from NNLO+NNLL calculations.

Diboson production is modelled using {\normalfont \scshape Sherpa}~\cite{Gleisberg:2008ta} 
with the {\normalfont \scshape CT10} PDF set, 
and the yields are normalised to the NLO cross-sections: 23 pb 
($WW \rightarrow \ell \nu qq$), 
0.7 pb ($ZZ \rightarrow \ell \ell qq$), 6.0 pb ($WZ \rightarrow \ell \nu qq$) and 4.6 pb ($ZW \rightarrow \ell \ell qq$).  

\section{Object definitions}
\label{sec:obj}

Electron candidates are reconstructed using charged-particle tracks in 
the inner detector associated with energy deposits in the electromagnetic
calorimeter. Muon candidates are identified by matching track segments 
in the muon spectrometer with tracks in the inner detector.
Lepton candidates are required to be isolated using the
``mini-isolation'' criteria described in Ref.~\cite{Rehermann:2010vq}.

Jets are reconstructed using the anti-$k_t$ algorithm
\cite{Cacciari:2008gp} implemented in the {\sc FastJet} package
\cite{Cacciari:2011ma} with radius 
parameter $R =$ 0.4 (small-$R$) or $R =$ 1.0 (large-$R$), using as input calibrated topological
clusters~\cite{Aad:2011he} of energy deposits in the calorimeters.
The jet-trimming algorithm \cite{Krohn:2009th} is applied to the large-$R$ jets 
to reduce the effect of soft and diffuse radiation, such as that from pile-up, multiple parton interactions and initial-state radiation. 
Large-$R$ jets are trimmed by reclustering the constituents with the $k_t$ algorithm~\cite{Catani:1993hr,Ellis:1993tq} with a 
radius parameter $R_{\text{sub}} =$ 0.3 and retaining sub-jets that have a momentum exceeding 5\% of that of the large-$R$ jet ($f_{\text{cut}}=$ 0.05).  
For small-$R$ jets, a pile-up correction based on the jet area, the
number of primary vertices, the bunch spacing, and jet \eta{} is applied.  Both jet collections are 
calibrated to the stable-particle level as a function of \pt{} and \eta{} (and mass for large-$R$ jets)~\cite{Aad:2013gja}. The stable-particle level 
refers to generator-level jets reconstructed from particles
with a lifetime of at least 10 ps. Small-$R$ jets are $b$-tagged using an algorithm that exploits the relatively 
large decay time of $b$-hadrons and their large mass~\cite{BTaggingEfficiency,CLTaggingEfficiency}.

The missing transverse momentum (with magnitude $\met$) is computed as the negative vector sum of the energy of 
all calorimeter cells, taking into account the calibration of reconstructed objects,
and the presence of muons.

\section{Event selection and reconstruction}
\label{sec:evtselreco}

Each event must have a reconstructed primary vertex with five or more 
associated tracks of $\pt >$ 400 $\MeV{}$. The events are required to contain exactly one 
reconstructed lepton candidate, which must then be geometrically matched to the trigger object.
To reduce the multi-jet background, the magnitude of the missing transverse momentum and the $W$-boson transverse mass $m_{\mathrm{T}}^W$ must satisfy
$\met > 20$~\GeV and $E_{\mathrm{T}}^{\mathrm{miss}}~+~m_{\mathrm{T}}^W~>~60$~\GeV, where 

\begin{equation}
m_{\mathrm{T}}^W = \sqrt{ 2 p_{\mathrm{T}}^{\mathrm{lepton}} \met (1 - \cos \Delta\phi) }
\end{equation} 

and $\Delta\phi$ is the azimuthal angle between the lepton and the missing
transverse momentum. 
At least one small-$R$ jet ($R=$ 0.4) must be found close to, but not 
coincident with, the lepton 
($\Delta R(\ell,\mathrm{jet}_{R=0.4}) < 1.5$). 

The leptonic top-quark candidate
is reconstructed by adding the highest-\pt{} jet among those satisfying the
above criteria, the selected charged lepton and the reconstructed neutrino.
The longitudinal component of the neutrino 
momentum is calculated by constraining the lepton-plus-missing-momentum 
system to have the $W$ boson mass and solving the resulting 
quadratic equation. If two real solutions are found, the one 
that yields the smallest longitudinal momentum for the neutrino is used. 
If no real solution exists, the missing transverse momentum
vector is varied by the minimal amount required to produce
exactly one real solution.

The hadronically decaying top quark is reconstructed as a single trimmed
jet with $R=$ 1.0.  
The selected jet must have \pt{} $>$ 300~\gev, must 
be well separated from both the charged lepton
($\Delta \phi(\ell,\mathrm{jet}_{R=1.0}) > 2.3$) and the small-$R$ jet 
associated with the leptonic top-quark candidate 
($\Delta R(\mathrm{jet}_{R=1.0},\mathrm{jet}_{R=0.4}) > 1.5$).
A substructure analysis of the large-$R$ jet is used to tag the
boosted top-quark candidate: 
the invariant mass of the jet $m_\mathrm{jet}^{\rm{trim}}$ after calibration
to the particle level~\cite{ATLAS:2012am} must be larger than
100~\GeV{} and the $k_t$ splitting scale\footnote{The $k_t$ splitting 
scale~\cite{Butterworth:2002tt} is obtained by reclustering the 
large-$R$ jet components 
with the $k_t$ algorithm with a radius parameter $R =$ 0.3. The first
splitting scale $\sqrt{d_{12}^{\rm{trim}} }$ corresponds 
to the scale at which the last two sub-jets are merged into one: $\sqrt{d_{12}^{\rm{trim}} } = \text{min}(p_{\text{T},1}, p_{\text{T},2}) \times \Delta R_{1,2}$, where 1 and 2 denote the two sub-jets merged in the last step of the $k_t$ algorithm.} 
$\sqrt{d_{12}^{\rm{trim}} }$ must exceed 40~\gev. 

Finally, at least one of the highest-$\pt$ small-$R$ jets associated with 
the decay of a top-quark candidates ($\Delta R(\ell,\mathrm{jet}_{R=0.4}) < 1.5$ or 
$\Delta R(\mathrm{jet}_{R=1.0},\mathrm{jet}_{R=0.4}) < 1.0$) must be $b$-tagged.
Events with a reconstructed \ttbar{} mass of less than 750~\gev{} are 
rejected, as the performance of the reconstruction of boosted top quarks 
is strongly degraded at low mass. 

The selection and reconstruction schemes yield good efficiency
and $\ttbar{}$ mass determination for high-mass pairs. 
Detailed MC studies presented in Ref.~\cite{Aad:2015fna} show that the 
mass resolution is approximately 6\% for a large range of \ttbar{} mass, 
starting at $m_{\ttbar} \sim$ 1~\TeV. The measurement of the top and 
anti-top-quark rapidities are nearly unambiguous. The quality of the top 
quark rapidity reconstruction can be expessed in
terms of the dilution factor $ D = 2p - 1$, where $p$ is the probability 
of a correct assignment of the $\Delta |y|$ sign. A dilution 
factor $D=$ 1 indicates perfect charge assignment.
The MC simulation predicts a value of approximately $D=$ 0.75 
for the selected sample. The remaining dilution is largely due to events with 
small values of the absolute rapidity difference; if events with
$|\Delta |y|| <$ 0.5 are excluded, the MC simulation predicts a
dilution factor greater than 0.9. 

\section{Comparison of data to the SM template}
\label{sec:yield}

A template for the expected yield of most SM processes is based on 
Monte Carlo simulation, where the production rate is normalised using 
the prediction of the inclusive cross-section 
specified in Section~\ref{sec:mc}. 
Exceptions are the $W$+jets background and 
the multi-jet background. The $W$+jets background 
normalisation and heavy-flavour fractions are corrected with scale factors derived from data, 
as in Ref.~\cite{Aad:2015fna}, using the observed asymmetry in the yields of 
positively and negatively charged leptons. The multi-jet background estimate is fully data-driven, using the matrix method. This method 
uses the selection efficiencies of leptons from prompt and non-prompt sources to predict the number of events with non-prompt leptons in the signal region.
These methods and their results are documented in detail in 
Ref.~\cite{Aad:2015fna}.

The event yields are compared to the SM expectation in 
Table~\ref{tab:yield}. The distributions of two key observables, 
the invariant mass of the \ttbar{} 
system and the difference of the absolute rapidities of the candidate top and 
anti-top quarks are shown in Figure~\ref{fig:detector_level}, for the combination 
of the \ejets{} and \mujets{} channels.
The observed event yield is approximately 10\% less than the MC prediction, 
the result of the softer top-quark \pt{} spectrum in data, which is also 
reported in Refs.~\cite{Aad:2015hna,Aad:2014zka,Khachatryan:2015oqa}.

\begin{table}[tbh!]
\centering
\caption{Observed and expected number of events in the signal region. The two columns correspond to the 
$e$+jets and $\mu$+jets selected data samples. The systematic uncertainties of the SM expectation 
include those from detector-related uncertainties, uncertainties in the normalisation, the 
luminosity uncertainty and the uncertainty in the cross-section predictions used to normalise the expected yields. }
\begin{tabular}{lr@{ $\pm$ }lr@{ $\pm$ }l}
\toprule
              & \multicolumn{2}{c}{\ejets{}}    & \multicolumn{2}{c}{\mujets{}}  \\
\midrule
$\ttbar$      & 4100 & 600 & 3600 & 500 \\
\wjets{}      & 263  & 32  & 264 & 32  \\
Single top    & 140  & 20  & 138  & 19  \\
Multi-jet      & 44  & 8   & 4 & 1   \\
\zjets{}      & 40   & 27  & 16 & 11  \\
Dibosons      & 20   & 7   & 18 & 7   \\
\ttV{}        & 37   & 19  & 33 & 17  \\
\midrule
Prediction    & 4600 & 600 & 4100 & 500 \\
\midrule
Data          & \multicolumn{2}{c}{4141}       & \multicolumn{2}{c}{3600}           \\
\bottomrule
\end{tabular}

\label{tab:yield}
\end{table}

Since $A_{\text{C}}$ is measured as a ratio, it is not sensitive to the 
absolute cross-section. The impact of the differences in the expected and 
observed shapes of the distributions in Figure~\ref{fig:detector_level} on the 
measurement is estimated by reweighting the simulated $\Delta|y|$ and 
top quark \pt{} distributions to match the data and found to be negligible.

\begin{figure} 
\centering
 \subfigure[]{\includegraphics[width=0.45\textwidth]{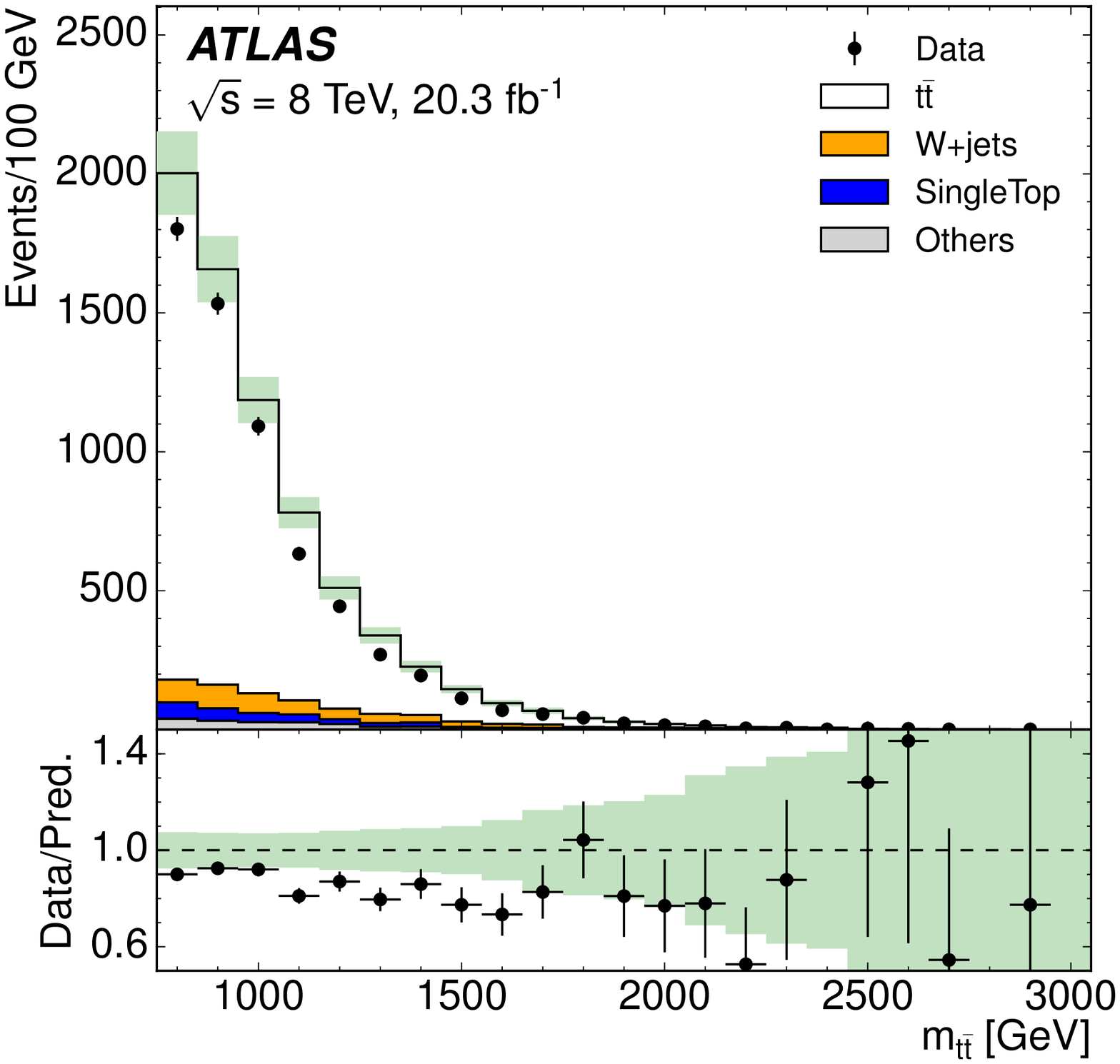}}
 \subfigure[]{\includegraphics[width=0.45\textwidth]{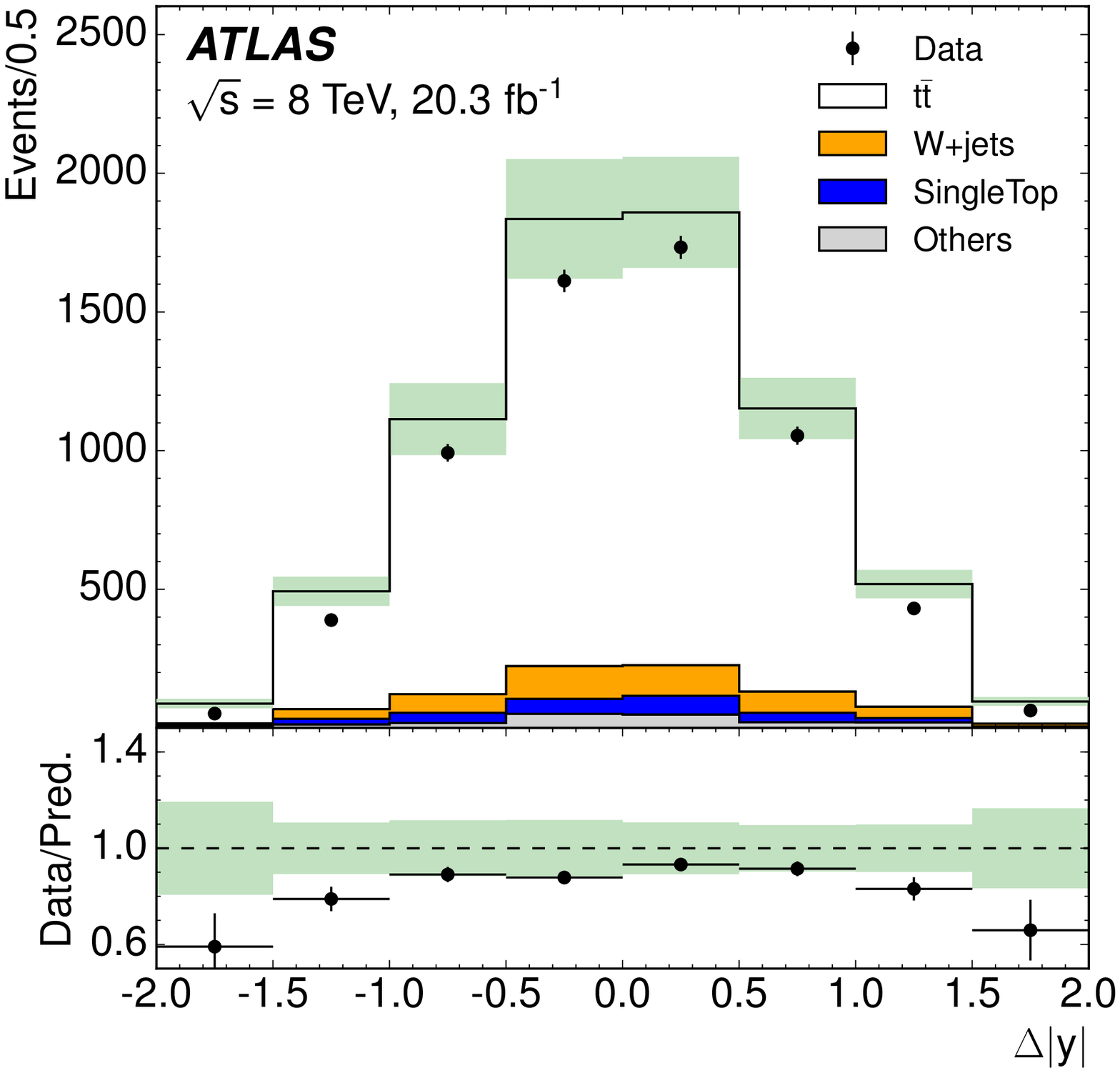}}
\caption{Detector-level distributions of (a) the invariant mass of the \ttbar{} system and (b) the difference of the absolute rapidities $\Delta|y|$ of top and anti-top-quark candidates, for the combination of the \ejets{} and \mujets{} channels. The observed distributions are compared to the SM expectation based on a mixture of data-driven techniques and Monte Carlo simulation. The ratio of data to the SM expectation is shown in the lower plots. Error bars on the data points indicate the statistical uncertainty. The hashed area shows the uncertainty of the SM prediction. This includes the statistical uncertainty, the theory uncertainties in the cross-sections, the effect of detector systematic uncertainties on the expected yield, the luminosity uncertainty, the uncertainty in the normalisations, and the signal modelling uncertainty. }
\label{fig:detector_level}
\end{figure}

\section{Unfolding}
\label{sec:unfolding}

An unfolding procedure transforms the observed charge asymmetry into a 
parton-level result in the phase space covered by the measurement:

\begin{equation}
m_{\ttbar} > 750 \text{\GeV,}  -2 < \Delta |y| < 2.
\label{eq:phasespace}
\end{equation}

The corrected result can thus be compared directly to fixed-order calculations
that implement these constraints.

The unfolding procedure is identical to the one used in a previous ATLAS charge
asymmetry measurent~\cite{ATLAS-TOPQ-2014-16}.
The \eplus\ and \muplus\ channels are combined to form a single set of events.
The data are corrected for migrations due to detector resolution using
a matrix unfolding method based on the open 
source \href{https://pypi.python.org/pypi/fbu}{PyFBU} implementation 
of the fully Bayesian unfolding (FBU)~\cite{Choudalakis:2012hz} algorithm. 
A bias in the charge asymmetry introduced by the selection criteria is 
corrected using a bin-by-bin acceptance correction.

The asymmetry in the full region of Eq.~(\ref{eq:phasespace}) is obtained
by correcting the content of four $\Delta |y|$ bins 
with the following boundaries: [$-2$, $-0.7$, 0, 0.7, 2].  For simulated 
events with a reconstructed $\Delta |y|$ that falls within $-2 < \Delta |y| < 2$, 
but a true $\Delta |y|$ outside this boundary (0.1\% of events),
the true value is included in the outermost $\Delta |y|$ bin. 
A differential result in three $m_{\ttbar}$ intervals (0.75~\TeV $< m_{\ttbar} <$ 0.9~\TeV, 
0.9~\TeV $< m_{\ttbar} <$ 1.3~\TeV, and 1.3~\TeV $< m_{\ttbar}$) is obtained using a
(1+12) $\times$ 12 matrix that corrects for mass and $\Delta |y|$ migrations.
The extra {\em underflow} bin keeps track of migrations of selected events from 
outside of the fiducial volume, $m_{\ttbar} <$ 0.75~\tev.
The $\Delta |y|$ binning in each mass bin is optimised to yield minimal
bias when non-SM asymmetries are injected. 

Uncertainties due to limitations in the understanding of object reconstruction and in the calibration
of the experiment described in Section~\ref{sec:syst} 
are included as nuisance parameters in the unfolding procedure, as well as 
the normalisation of the backgrounds. In this study, the data sample is 
too small for FBU to significantly constrain any of the nuisance parameters, and therefore the size of the detector-related and 
normalisation uncertainties are not reduced by the unfolding process.

\section{Systematic uncertainties}
\label{sec:syst}

Systematic uncertainties are estimated as in Ref.~\cite{Aad:2015fna} 
and propagated to the $A_{\mathrm{C}}$ measurement following the procedure of
Ref.~\cite{ATLAS-TOPQ-2014-16}. The non-negligible uncertainties in the unfolded charge asymmetry 
measurement are presented in Table~\ref{tab:systematics_table}.

The most important uncertainties among the detector-related and background 
normalisation uncertainties are the 
scale and resolution of the jet energy (17 nuisance parameters for large-$R$ jets
and 21 for small-$R$ jets) and the $b$-tagging performance (10 nuisance 
parameters)~\cite{Aad:2014bia,Aad:2011he,Aad:2012vm}. The impact of uncertainties in the reconstruction of
electrons and muons and the missing transverse momentum is 
negligible. Detector-related uncertainties and background normalisation uncertainties
have a small impact on the analysis.

The uncertainty due to imperfections in the MC generator modelling is estimated
using a number of alternative generators. The most important effects are
the choice of NLO ME and parton shower/hadronisation model. Each alternative sample is unfolded 
using the nominal procedure. The ME modelling uncertainty is taken as the difference 
between the results for {\normalfont \scshape Powheg-Box}$+${\normalfont \scshape Herwig}/{\normalfont \scshape Jimmy} and 
{\normalfont \scshape MC@NLO}$+${\normalfont \scshape Herwig}/{\normalfont \scshape Jimmy}. The PS/hadronisation modelling uncertainty is evaluated
as the difference between {\normalfont \scshape Powheg-Box}$+${\normalfont \scshape Pythia} and 
{\normalfont \scshape Powheg-Box}$+${\normalfont \scshape Herwig}/{\normalfont \scshape Jimmy}. The results are corrected for the
small differences in the prediction of the true $A_{\mathrm{C}}$ among the
generators. The ISR/FSR uncertainty is estimated 
as half the difference between two {\normalfont \scshape AcerMC} samples 
with radiation settings varied within the range allowed by data. The uncertainty associated with the choice of 
PDF is evaluated using the {\normalfont \scshape MC@NLO}$+${\normalfont \scshape Herwig}/{\normalfont \scshape Jimmy} sample, 
by comparing the differences when reweighting the sample to CT10, MSTW 2008~\cite{mstw}, and NNPDF2.1~\cite{nnpdf} PDF sets.
The three contributions are assumed to be uncorrelated and are added in quadrature, 
forming the dominant systematic uncertainty in the measurement.

\begin{table}[h!]
  \begin{center}  
 \caption[]{The effect on the corrected charge asymmetry, in each $m_{\ttbar} $ interval, of systematic uncertainties on the signal and background modelling and the description of the detector response. The 
uncertainties are given in absolute percentages.}
    \begin{tabular}{l r r r r} % lcr->LeftCenterRight
\hline 	
 
$ m_{\ttbar} $ interval            &   $ > $ 0.75~\tev{}    & 0.75 -- 0.9~\tev{}    & 0.9  -- 1.3~\tev{}  & $>$ 1.3~\tev{} \\ 
\hline 

      \multicolumn{5}{c}{Breakdown of detector-related systematic uncertainties} \\ 
     \hline
      Jet energy and resolution - $R=$ 0.4 jets	&  0.1\%   &  0.4\% & 0.3\% & 0.4\% \\
      Jet energy and resolution - $R=$ 1.0 jets	& 0.3\%  & 1.6\%   & 0.6\% & 1.0\% \\
      $b$-tag/mis-tag efficiency 		        & 0.2\%  & 0.2\% & 0.2\% & 0.7\% \\
      Lepton reconstruction/identification/scale 			& 0.1\% & 0.2\%  & 0.1\% & 0.1\%  \\
      Missing transverse momentum (\met) 		& 0.1\%	& $<$ 0.1\%   & $<$ 0.1\% & 0.1\%  \\
      Background normalization  		& 0.1\%		&  0.2\%  & 0.3\% & 0.4\%  \\  \hline
      \multicolumn{5}{c}{Combined detector-related uncertainties and others} \\ 
      \hline
      Statistical + detector-related systematic                    & 2.0\%  & 6.0\% & 4.1\% & 11.6\% \\
      Signal modelling - matrix element     &  1.5\%   & 2.4\% &  0.6\%  & 5.3\%    \\
      Signal modelling - parton shower     &  2.0\%   & 3.2\% &  1.2\%  &  6.2\%   \\
      Signal modelling - ISR/FSR & 0.1\%  &  0.3\% & 0.1\%  & 3.0\% \\ 
      Signal modelling - PDF & 0.4\%  &  0.4\% & 0.3\%  & 3.3\% \\ 
      Unfolding \& MC statistics 	& 0.5\%  & 1.2\% & 0.8\% & 2.1\% \\ \hline 
      Total                     & 3.2\%  & 7.3\% & 4.4\% & 15.0\% \\ \hline             
    \end{tabular}
 
    \label{tab:systematics_table}
  \end{center}
\end{table}

The unfolding uncertainty includes two components. The first component, 
the uncertainty due to the limited number of events in the Monte Carlo 
samples used to correct the data, is estimated by propagating the statistical 
uncertainty of the elements of the response matrix with pseudo-experiments.
To evaluate the second component due to the non-linearity of the unfolding, 
different charge asymmetry values are injected by reweighting the \ttbar{} 
Monte Carlo sample according to several functional forms. 
The uncertainty is taken as the bias estimated for the observed charge 
asymmetry values. A number of stress tests are
performed, where the MC samples are reweighted to mimic the observed
differences in the $m_{\ttbar}$ and $\Delta|y|$ distributions. The impact
on the results of the unfolding procedure is found to be small compared to the
unfolding uncertainty and is not taken into account as a separate 
uncertainty. In addition, the measurement is performed in a 
more restricted $|\Delta |y||$ region, excluding events with $|\Delta |y|| <$ 0.5, 
where the dilution factor $D$ is smaller. The result is found to be consistent 
with the nominal measurement, and no uncertainty 
is assigned.

\begin{figure}[htb]
\begin{center}
\includegraphics[width=0.7\textwidth]{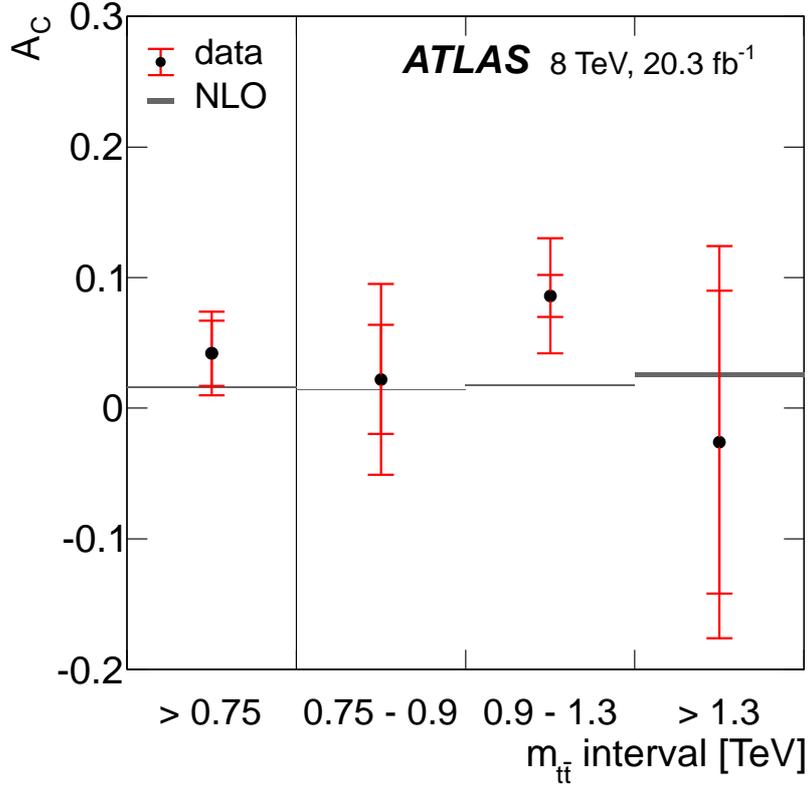}
\caption{A summary of the charge asymmetry measurements.  The error bars on the data indicate the modelling 
and unfolding systematic uncertainties, shown as the inner bar, and the total uncertainty, which includes the statistical 
uncertainty and the experimental systematic uncertainties. The SM prediction of the NLO calculation in Ref.~\cite{Kuhn:2011ri} 
for the charge asymmetry of top-quark pairs with $|\Delta |y|| <$ 2 is indicated as a shaded horizontal bar in each $m_{\ttbar{}}$ bin, 
where the width of the bar indicates the uncertainty.}
\label{fig:summary_all}
\end{center}
\end{figure}

\section{Results}
\label{sec:results}

The results for the charge asymmetry in the four $m_{\ttbar}$ intervals 
are presented in Figure~\ref{fig:summary_all} and Table~\ref{tab:results}. 
The measurement for $m_{\ttbar}$ > 0.75~\tev{} and $|\Delta|y||<$ 2
yields $A_{\mathrm{C}} = $ (4.2 $\pm$ 3.2)\%, where the uncertainty is dominated
by the modelling uncertainty, followed by the statistical uncertainty 
of the data. The result is within one standard deviation of the SM expectation.
A differential measurement is also presented, in three $m_{\ttbar}$ bins: 0.75 -- 0.9~\tev{}, 
0.9 -- 1.3~\tev{} and $m_{\ttbar} >$ 1.3~$\TeV$ ($|\Delta|y||<$ 2 for all measurements). The largest difference with respect to the 
SM prediction is observed in the bin with $m_{\ttbar}$ = 0.9 -- 1.3~\TeV, where it reaches 1.6$\sigma$. 

\begin{table}[tbh!]
\centering
\caption{The measured charge asymmetry after the unfolding to parton level in four intervals of the 
invariant mass of the \ttbar{} system. The result is compared to the SM prediction using the NLO 
calculation in Ref.~\cite{Kuhn:2011ri}. The phase space is limited to $|(\Delta |y|)|<$ 2. The uncertainties correspond to the sum in quadrature of statistical and systematic uncertainties (for the data) or to the theory uncertainty (for the SM prediction).  }
\label{tab:results}
\begin{tabular}{lr@{ $\pm$ }lr@{ $\pm$ }lr@{ $\pm$ }lr@{ $\pm$ }l}
\toprule

  $ m_{\ttbar} $ interval            &   \multicolumn{2}{c}{$ > $ 0.75~\tev}    &  \multicolumn{2}{c}{0.75 -- 0.9~\tev}    &  \multicolumn{2}{c}{0.9  -- 1.3~\tev}  &  \multicolumn{2}{c}{$>$ 1.3~\tev} \\
\hline
Measurement   &     (4.2 & 3.2)\%   &     (2.2 & 7.3)\%            &     (8.6 & 4.4)\%          &     ($-$2.9 & 15.0) \%          \\
SM prediction &        (1.60 & 0.04)\% &        (1.42 & 0.04)\% & (1.75 & 0.05)\% & (2.55 & 0.18)\% \\
\bottomrule
\end{tabular}
\end{table}

\section{Impact on BSM scenarios}
\label{sec:bsm}

\begin{figure}[htbp!]
\begin{center}
\subfigure[]{\includegraphics[width=0.6\textwidth,clip]{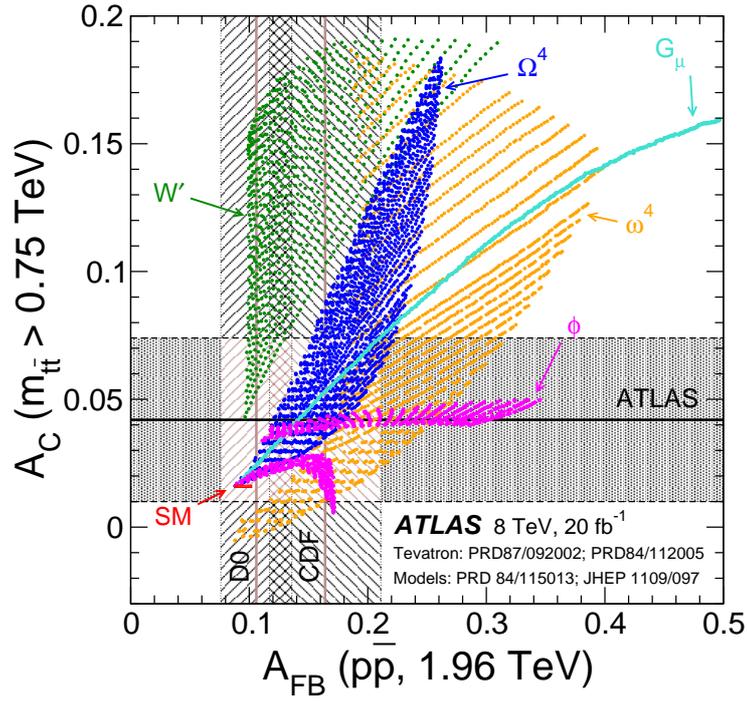}}
\subfigure[]{\includegraphics[width=0.6\textwidth,clip]{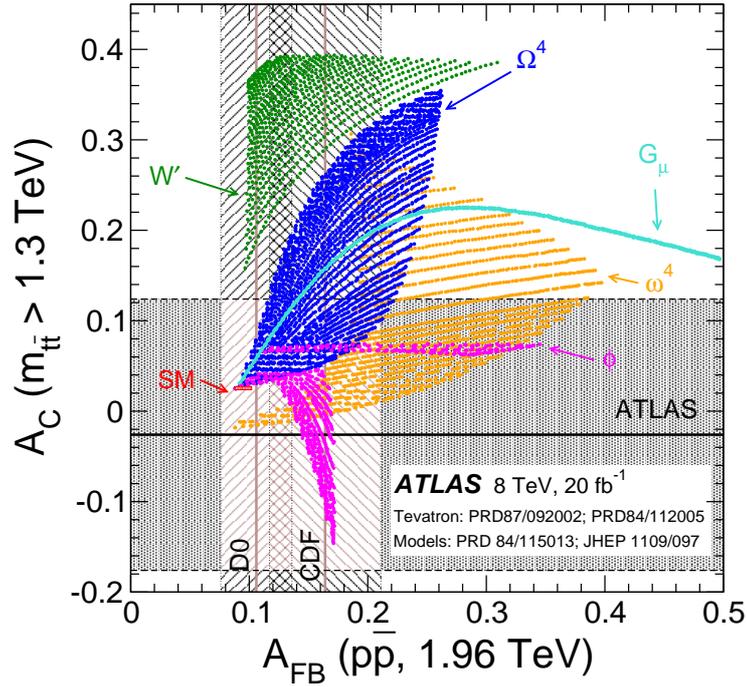}}
\caption{Predictions from a number of extensions of the SM from Refs.~\cite{AguilarSaavedra:2011hz,AguilarSaavedra:2011ug}, for the forward--backward 
asymmetry integrated over $m_{\ttbar}$ at the Tevatron (on the $x$-axis in both plots) and two high-mass charge asymmetry measurements at the LHC. 
The $y$-axis in both figures represents the measurement for (a) $m_{\ttbar} > $ 0.75~\tev{} and for (b) $m_{\ttbar} > $ 1.3~\tev{}. The SM 
predictions of both the forward--backward asymmetry at the Tevatron and the charge asymmetry at the LHC are also shown~\cite{Czakon:2014xsa,Bern:2012}.}
\label{fig:impact_model_zoo}
\end{center}
\end{figure}

Extensions of the SM with heavy particles can predict a significantly enhanced 
high-mass charge asymmetry at the LHC. In Figure~\ref{fig:impact_model_zoo},  
BSM predictions of the charge asymmetry in 8~\tev{} $pp$ collisions
with $m_{\ttbar} > $ 0.75~\tev{} and $m_{\ttbar} > $ 1.3~\tev{} 
are compared with $A_{\mathrm{FB}}$ integrated over $m_{\ttbar}$ in $p\bar{p}$ collisions at $\sqrt{s}=$ 1.96~\tev{}. 
The measurements presented in this Letter are indicated as horizontal bands.
The measurements of $A_{\mathrm{FB}}$ integrated over $m_{\ttbar}$ in top-quark
pair production at 1.96~\tev{} in $p\bar{p}$ collisions by CDF~\cite{Aaltonen:2012it} 
and D0~\cite{Abazov:2011rq} are shown as vertical bands.

The clouds of points in Figure~\ref{fig:impact_model_zoo} correspond to a number of models in 
Refs.~\cite{AguilarSaavedra:2011hz,AguilarSaavedra:2011ug}: 
a heavy $W'$ boson exchanged in the $t$-channel, a heavy axi-gluon 
$G_{\mu}$ exchanged in the $s$-channel, and doublet ($\phi$), triplet ($\omega^4$) 
or sextet ($\Omega^4$) scalars. Each point corresponds to a choice of the 
new particle's mass, in the range between 100~\gev{} and 10~\tev{}, and of
the couplings to SM particles, where all values allowed give a total cross-section for
top-quark pair production at the Tevatron compatible with observations
and a high-mass \ttbar{} production cross-section ($m_{\ttbar} >$ 1~\tev{}) at the LHC that 
is at most three times the SM prediction. The contribution 
from new physics to the Tevatron $A_{\mathrm{FB}}$ is moreover required to be
positive. The predictions of the Tevatron forward--backward asymmetry 
and the LHC high-mass charge asymmetry are calculated using 
PROTOS~\cite{AguilarSaavedra:2008gt}, which includes the tree-level SM 
amplitude plus the one(s) from the new particle(s), taking into account 
the interference between the two contributions. 
This measurement extends the reach of previous ATLAS and CMS measurements to beyond 1~\TeV (adding a bin with $m_{\ttbar{}} =$ 0.9 - 1.3~\TeV{}). 
The BSM sensitivity of this measurement is also complementary to that of 
the most recently published ATLAS measurement~\cite{ATLAS-TOPQ-2014-16}
and can be seen to disfavour the $t$-channel $W'$ boson model 
in the highest $m_{\ttbar}$ bin. 

\section{Conclusions}
\label{sec:conclusions}

The charge asymmetry in the rapidity distribution of top-quark pairs produced at large \ttbar{} invariant
mass has been measured in a sample of $\sqrt{s} = $ 8~\tev{} $pp$ collisions corresponding 
to an integrated luminosity of 20.3~\ifb, collected with the ATLAS experiment 
at the LHC in 2012. The selection criteria and the reconstruction algorithm designed for \ljets events 
with the decay topology of highly boosted top quarks are found to give 
good control over the sign of the absolute rapidity
difference of top and anti-top quarks, with a dilution factor that 
reaches 0.75, significantly higher than more traditional methods. 

The observed asymmetry is corrected to the fiducial space $m_{\ttbar} > $ 0.75~\tev{} 
and $-2  < \Delta |y| < 2$. The result, $A_{\mathrm{C}} = $ (4.2 $\pm$ 3.2)\%, is less than 
one standard deviation from the SM prediction of 1.60 $\pm$ 0.04 \%. The charge asymmetry is 
also determined in three \ttbar{} mass intervals.
The most significant deviation from the SM prediction, 1.6$\sigma$, is observed in the 
mass bin that ranges from 0.9~\tev{} to 1.3~\tev: $A_{\mathrm{C}} =$ (8.6 $\pm$ 4.4)\%.
The other two mass bins yield values compatible with the SM 
prediction within 1$\sigma$. These measurements provide a constraint on extensions of the SM, some of which predict a very sizeable charge asymmetry at large \ttbar{} mass.

\section*{Acknowledgements}

% Acknowledgements for papers with collision data
% Version 15-Sep-2015

% Standard acknowledgements start here
%----------------------------------------------
We thank CERN for the very successful operation of the LHC, as well as the
support staff from our institutions without whom ATLAS could not be
operated efficiently.

We acknowledge the support of ANPCyT, Argentina; YerPhI, Armenia; ARC, Australia; BMWFW and FWF, Austria; ANAS, Azerbaijan; SSTC, Belarus; CNPq and FAPESP, Brazil; NSERC, NRC and CFI, Canada; CERN; CONICYT, Chile; CAS, MOST and NSFC, China; COLCIENCIAS, Colombia; MSMT CR, MPO CR and VSC CR, Czech Republic; DNRF, DNSRC and Lundbeck Foundation, Denmark; IN2P3-CNRS, CEA-DSM/IRFU, France; GNSF, Georgia; BMBF, HGF, and MPG, Germany; GSRT, Greece; RGC, Hong Kong SAR, China; ISF, I-CORE and Benoziyo Center, Israel; INFN, Italy; MEXT and JSPS, Japan; CNRST, Morocco; FOM and NWO, Netherlands; RCN, Norway; MNiSW and NCN, Poland; FCT, Portugal; MNE/IFA, Romania; MES of Russia and NRC KI, Russian Federation; JINR; MESTD, Serbia; MSSR, Slovakia; ARRS and MIZ\v{S}, Slovenia; DST/NRF, South Africa; MINECO, Spain; SRC and Wallenberg Foundation, Sweden; SERI, SNSF and Cantons of Bern and Geneva, Switzerland; MOST, Taiwan; TAEK, Turkey; STFC, United Kingdom; DOE and NSF, United States of America. In addition, individual groups and members have received support from BCKDF, the Canada Council, CANARIE, CRC, Compute Canada, FQRNT, and the Ontario Innovation Trust, Canada; EPLANET, ERC, FP7, Horizon 2020 and Marie Skłodowska-Curie Actions, European Union; Investissements d'Avenir Labex and Idex, ANR, Region Auvergne and Fondation Partager le Savoir, France; DFG and AvH Foundation, Germany; Herakleitos, Thales and Aristeia programmes co-financed by EU-ESF and the Greek NSRF; BSF, GIF and Minerva, Israel; BRF, Norway; the Royal Society and Leverhulme Trust, United Kingdom.

The crucial computing support from all WLCG partners is acknowledged
gratefully, in particular from CERN and the ATLAS Tier-1 facilities at
TRIUMF (Canada), NDGF (Denmark, Norway, Sweden), CC-IN2P3 (France),
KIT/GridKA (Germany), INFN-CNAF (Italy), NL-T1 (Netherlands), PIC (Spain),
ASGC (Taiwan), RAL (UK) and BNL (USA) and in the Tier-2 facilities
worldwide.
%----------------------------------------------

\clearpage
\appendix

\printbibliography

\newpage % ATLAS Collaboration author list
% Data extracted on 18-Dec-2015 for paper reference TOPQ-2015-10

\begin{flushleft}
{\Large The ATLAS Collaboration}

\bigskip

G.~Aad$^\textrm{\scriptsize 87}$,
B.~Abbott$^\textrm{\scriptsize 114}$,
J.~Abdallah$^\textrm{\scriptsize 152}$,
O.~Abdinov$^\textrm{\scriptsize 11}$,
B.~Abeloos$^\textrm{\scriptsize 118}$,
R.~Aben$^\textrm{\scriptsize 108}$,
M.~Abolins$^\textrm{\scriptsize 92}$,
O.S.~AbouZeid$^\textrm{\scriptsize 138}$,
H.~Abramowicz$^\textrm{\scriptsize 154}$,
H.~Abreu$^\textrm{\scriptsize 153}$
% ...
\bigskip
\\
% ...
$^{11}$ Institute of Physics, Azerbaijan Academy of Sciences, Baku, Azerbaijan\\
% ...
$^{87}$ CPPM, Aix-Marseille Universit{\'e} and CNRS/IN2P3, Marseille, France\\
% ...
$^{92}$ Department of Physics and Astronomy, Michigan State University, East Lansing MI, United States of America\\
% ...
$^{108}$ Nikhef National Institute for Subatomic Physics and University of Amsterdam, Amsterdam, Netherlands\\
% ...
$^{114}$ Homer L. Dodge Department of Physics and Astronomy, University of Oklahoma, Norman OK, United States of America\\
% ...
$^{118}$ LAL, Univ. Paris-Sud, CNRS/IN2P3, Universit{\'e} Paris-Saclay, Orsay, France\\
% ...
$^{138}$ Santa Cruz Institute for Particle Physics, University of California Santa Cruz, Santa Cruz CA, United States of America\\
% ...
$^{152}$ Institute of Physics, Academia Sinica, Taipei, Taiwan\\
$^{153}$ Department of Physics, Technion: Israel Institute of Technology, Haifa, Israel\\
$^{154}$ Raymond and Beverly Sackler School of Physics and Astronomy, Tel Aviv University, Tel Aviv, Israel\\
% ...
\end{flushleft}

% Created with xml2latex.py

\end{document}